\documentclass[10pt,sigconf,letterpaper,nonacm]{acmart}

\usepackage[english]{babel}
\usepackage{blindtext}
\usepackage{cleveref}
\usepackage{subcaption}
\usepackage{caption}
\usepackage{gensymb}

\providecommand{\ie}{\emph{i.e.,} }
\providecommand{\eg}{\emph{e.g.,} }

\providecommand{\myparab}[1]{\smallskip\noindent\textbf{#1} }

\newenvironment{packeditemize}{\begin{list}{$\bullet$}{\setlength{\itemsep}{0.2pt}\addtolength{\labelwidth}{0pt}\setlength{\leftmargin}{\labelwidth}\setlength{\listparindent}{\parindent}\setlength{\parsep}{1pt}\setlength{\topsep}{0pt}}}{\end{list}}
    
\usepackage{array,tabularx}

\newenvironment{conditions*}
  {\par\vspace{\abovedisplayskip}\noindent
   \tabularx{\columnwidth}{>{$}l<{$} @{}>{${}}c<{{}$}@{} >{\raggedright\arraybackslash}X}}
  {\endtabularx\par\vspace{\belowdisplayskip}}

\renewcommand\footnotetextcopyrightpermission[1]{} 
\setcopyright{none}

\acmDOI{}

\acmISBN{}

\acmPrice{}
\usepackage{caption}
\begin{document}

\title{Making Sense of Constellations}
\subtitle{Methodologies for Understanding Starlink's Scheduling Algorithms}

\author{Hammas Bin Tanveer}
\affiliation{%
  \institution{University of Iowa}
}
\author{Mike Puchol}
\affiliation{%
  \institution{Google X}
}
\author{Rachee Singh}
\affiliation{%
  \institution{Cornell University}
}
\author{Antonio Bianchi}
\affiliation{%
  \institution{Purdue University}
}
\author{Rishab Nithyanand}
\affiliation{%
  \institution{University of Iowa}
}

\renewcommand{\shortauthors}{Tanveer, et al.}

\maketitle

\section*{Abstract}
Starlink constellations are currently the largest LEO WAN and have seen
considerable interest from the research community.
%
%
In this paper, we use high-frequency and high-fidelity measurements to
uncover evidence of hierarchical traffic controllers in Starlink --- a global controller
which allocates satellites to terminals and an on-satellite controller that
schedules transmission of user flows.
We then devise a novel approach for identifying how satellites are allocated to
user terminals. 
Using data gathered with this approach, we measure the characteristics of the
global controller and identify the factors that influence the allocation of
satellites to terminals.
Finally, we use this data to build a model which approximates Starlink's global
scheduler. Our model is able to predict the characteristics of the satellite
allocated to a terminal at a specific location and time with reasonably high
accuracy and at a rate significantly higher than baseline.
%

\section{Introduction}\label{sec:introduction}


Low-earth orbit (LEO) satellite networks are expected to play 
an important role in achieving global broadband-like Internet 
connectivity since they enable low-latency, last-mile connectivity without heavy
infrastructure costs (\eg cell towers and fiber deployments),
Unfortunately, prior work has shown lackluster network performance
from Starlink-connected end-hosts, both via measurements~\cite{michel2022first} 
as well as simulations~\cite{bhosale2023characterization, kassing2020exploring}.
Despite these findings, researchers have been unable to propose and 
validate methods for improving the performance of the Starlink network. 
This is, in large part, due to the opacity of the network and 
its scheduling algorithms. 
Knowledge of algorithms responsible for
determining which satellites route traffic from specific
user terminal locations is key to engineering performance 
improvements for the network. 
In this paper, we address this gap in knowledge.
In particular, we empirically uncover the scheduling algorithms used by the Starlink
network by analyzing data from high-frequency measurements
from four Starlink terminals (deployed both in the US and the EU) to 
servers co-located at their corresponding Point-of-Presence. 

%
From our longitudinal and high-frequency measurements, we find that 
Starlink routes traffic from user terminals to their ground stations 
in a two-step process.
First, a {\em global network controller} allocates a satellite to
each user terminal based on a variety of factors including 
load, geospatial conditions, satellite charge, etc. Our experiments 
show that these allocations are made every 15-seconds, globally.
Second, a {\em local on-satellite controller} schedules flows 
from the user terminals assigned to it.
Taken together, these findings suggest that the hierarchical 
traffic engineering mechanisms, commonly deployed in terrestrial 
WANs \cite{b4andafter}, are also deployed by Starlink. 
We were able to confirm the validity of these major findings 
using recent FCC filings by SpaceX \cite{Starlink45:online}.


This work is the first to uncover hierarchical controllers
and the characteristics of traffic engineering in Starlink,
due to the following three key reasons. 
First, our high-frequency (millisecond granularity) measurements
allow us to observe signatures of traffic engineering 
(\eg abrupt latency changes) that cannot be observed with the 
coarse-grained measurements of prior work \cite{michel2022first}.
Second, by co-locating our destination server at the 
Starlink PoP, we minimize the influence of terrestrial latency 
on our measurements.
Finally, our novel methodology for identifying the satellite 
currently serving a given terminal allows us to obtain 
ground-truth regarding the traffic engineering decisions 
made by Starlink.
%

\myparab{Contributions.}
This paper makes four main contributions.

\begin{packeditemize}
    \item We use high-frequency measurements to show evidence of
        hierarchical traffic engineering on Starlink (\S\ref{sec:engineering}).

    \item We develop a novel technique for identifying the satellite that serves
        a user terminal (\S\ref{sec:identifying}).

    \item We uncover the characteristics and preferences of Starlink's global
        scheduler --- \ie the algorithm responsible for allocating satellites to
        specific user terminals (\S\ref{sec:understanding}).

    \item We develop an approximation of the Starlink global
        scheduler that can predict the satellites allocated to 
        user terminals with reasonably high accuracy (\S\ref{sec:model}).

\end{packeditemize}

\section{Background} \label{sec:background}



\begin{figure}
	\includegraphics[width=0.48\textwidth]{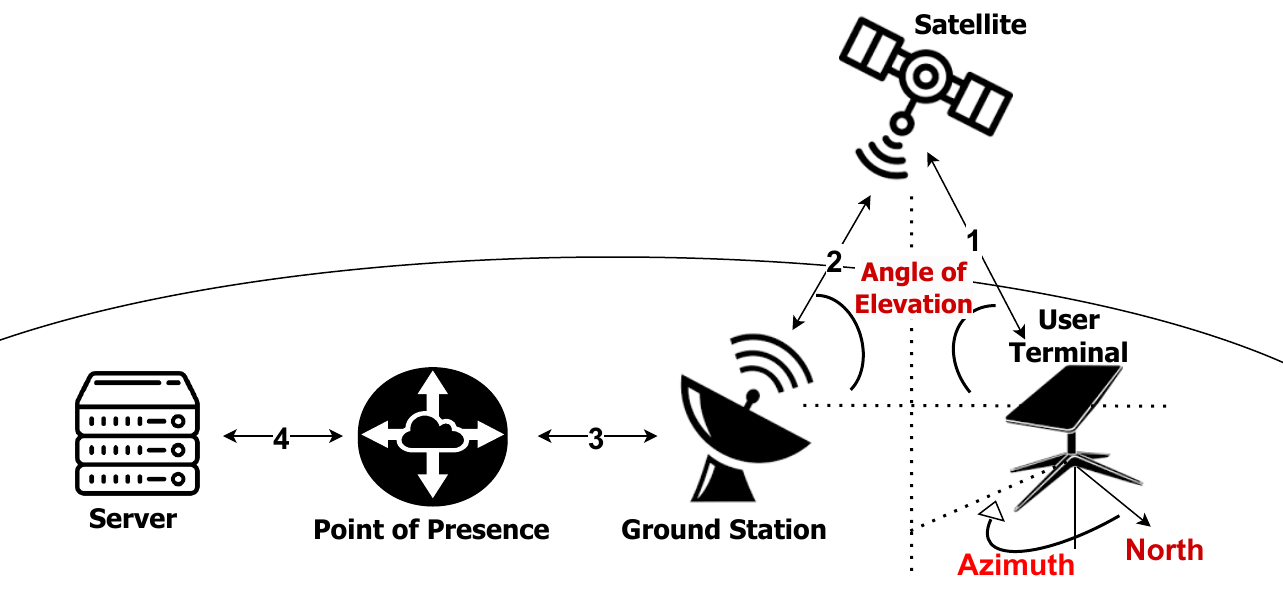}
        \caption{Starlink ecosystem.}
	\label{fig:msm}
\end{figure}

Starlink is a low-earth orbit satellite constellation consisting of nearly 4000
satellites in the low Earth orbit (LEO). 
%
The Starlink ecosystem has four key components: (1) In-orbit satellites, (2)
User terminals or dishes (3) Ground stations and (4) Points of Presence (PoPs).
\Cref{fig:msm} shows how these components interact with each other to provide
Internet connectivity to Satrlink's end users. 

\myparab{User terminals.}
Terminals are deployed on user premises to connect to in-orbit
satellites. Starlink user terminals are sophisticated phased-array antennas
equipped with a motor that can physically reposition the angle of the dish to
track fast moving satellites in the sky. 
User terminals can connect to any satellite at an angle of elevation higher than
25$^{\circ}$s (see \Cref{fig:msm}). While tens of satellites satisfy the angle
of elevation constraints, a terminal can connect to only one satellite at a
time. 
Terminals forward user traffic to the satellite assigned to them. Internals of
the algorithm that maps user terminals to satellites are currently known only to
the operators of the Starlink network. 
In this work, we empirically demonstrate characteristics of this algorithm and
build its approximation.

\myparab{Satellites.}
A Starlink satellite connects to multiple user terminals at a time. They
allocate radio frames to user terminals mapped to them to exchange data. 
In our work, we find evidence that this allocation is determined by a local
on-satellite controller.
In fact, we also find the description of this controller, referred to
as the medium access control scheduler~\cite{iyer2022system}, in recent FCC
filings from SpaceX. 
This controller considers factors such as user priority, current load, and
per-terminal flow characteristics when forwarding the traffic from user
terminals to ground stations. 

\myparab{Ground stations and PoPs.}
Ground stations consist of a set of phased-array antennas that receive traffic
from satellites and send it through wired links to Starlink's PoPs. 
Like user terminals, ground stations can communicate with satellites at an angle
of elevation higher than 25$^{\circ}$ above the horizon, relative to the ground
station.
A PoP is a terrestrial server with wired connectivity to a ground station. PoPs
are connected to the Internet backbone. From the PoP, the traffic is routed to
destination on the Internet.

\section{Evidence of Traffic Engineering}\label{sec:engineering}

\myparab{Experiment setup: Vantage points.}
We perform our measurement using four Starlink terminals --- one each in Western
Europe, Northeast US, Midwest US, and Northwest US.
To improve the precision of our measurements, we configured the Starlink router
to operate in bridge mode and connected them to a dedicated Raspberry Pi via
Ethernet. These Pi's were the source of our measurements. This approach prevents
the complexities which arise from using wireless routers in the measurement
infrastructure.
The destination of our measurements were servers co-located at the Starlink PoPs
assigned to the regions of our user terminals. This choice of destination allows
our measurements to be relatively unimpacted by the vagaries of terrestrial
networking.

\myparab{Experiment setup: Measurements.} We conduct high- frequency measurements
of the round-trip times and packet loss rates between our sources and
destination servers.
Packets were sent using iRTT \cite{irtt:online} at the rate of 1 packet/20 ms
and iPerf3 at a bandwidth of 50\% of the upstream connection. 
These parameters were chosen because they allowed stable and reliable
measurements of the Starlink network. At higher frequencies and bandwidths, the
packet loss rates and measured round-trip times were highly variable even within
the same measurement period.
From these measurements, we recorded high-resolution round-trip times and packet
loss rates. 
To facilitate accurate measurements of the round-trip times, the clocks of our
vantage points and servers were routinely synchronized using NTP.

\begin{figure}[th]
	\includegraphics[width=0.47\textwidth]{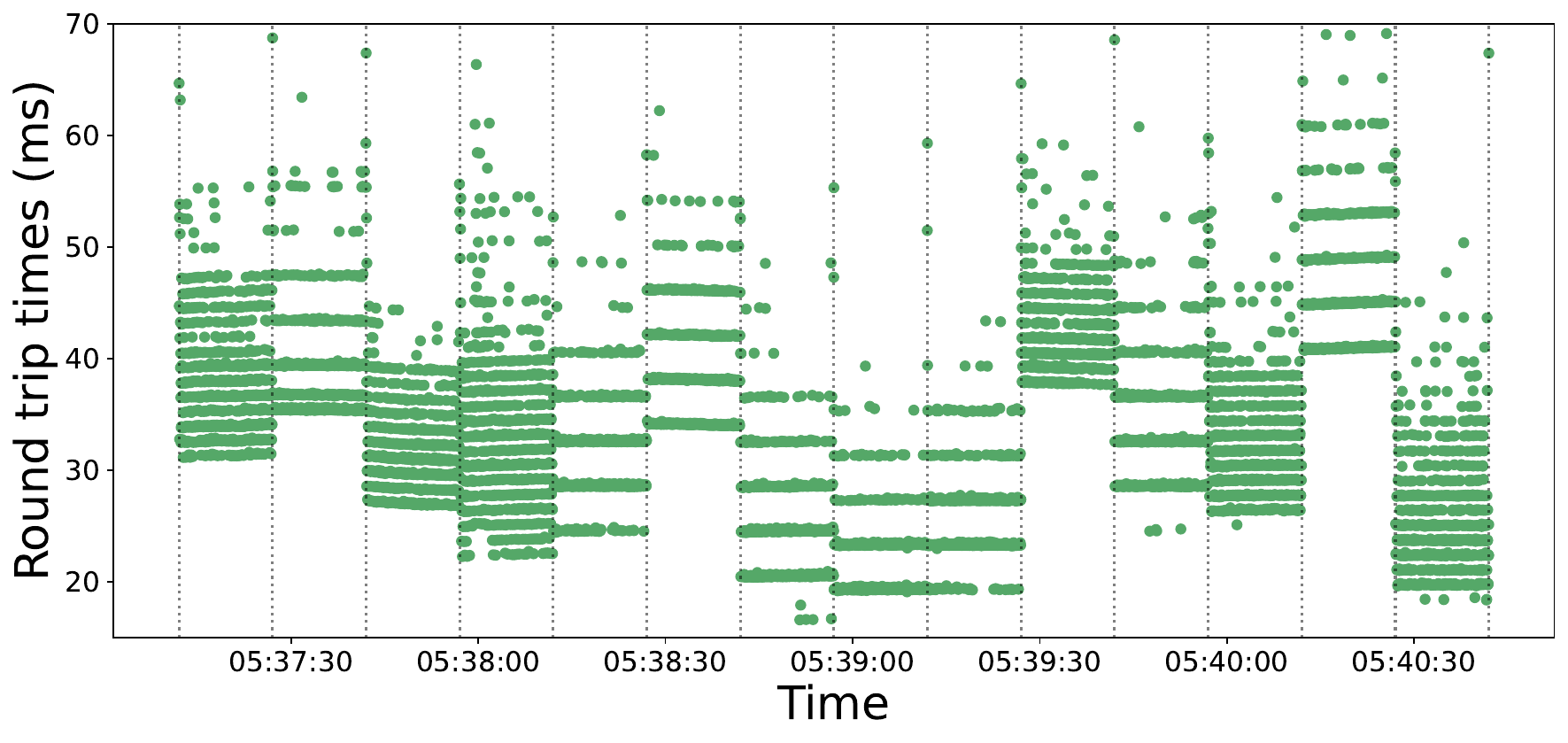}
	\caption{Measured RTT from EU terminal.}
	\label{fig:data:rtt}
\end{figure}

\myparab{Observation: Starlink relies on a global controller for
satellite-to-terminal scheduling.}
\Cref{fig:data:rtt} shows the changes in measured latency during a brief
measurement window of two minutes for our Midwest US terminal.
It is immediately obvious that major changes in latency characteristics
occur every 15 seconds --- specifically, at the 12th, 27th, 42nd, and 57th
second past every minute.
Notably, these changes are observed from all our measured locations 
for all periods of time.
In addition to visual tests, we are also able to confirm that the latency
characteristics observed during these consecutive 15-second windows are
statistically different (Mann-Whitney U test; $p$ < .05) from each other for all
locations and over the entire period of our measurements.
These drastic changes in latency are suggestive of global changes in the
satellites allocated to user terminals for several reasons, including:
(1) our measurements effectively nullify the impact of terrestrial networks; 
(2) these effects were observed, simultaneously, from all our vantage points;
and  
(3) these effects were noticed even when our terminals were running well under
capacity.
Upon further investigation, we discovered an FCC filing from SpaceX which
describes a global scheduler for periodically allocating terminals to satellites
\cite{Starlink45:online}. 
We conclude that our measurements have uncovered evidence of this scheduler and
the periodicity of reallocations.
It is important to note that this finding renders impossible the hypothesis that
performance characteristics are associated with the movement of the satellite
assigned to serve a terminal (\eg Figure 7 in \cite{kassem2022browser}). 
This is because changes in satellite allocation occur every 15 seconds which
is insufficient time to meaningfully cause impacts on performance due to change
in satellite positions/distances.
In the remainder of this paper, we will uncover the characteristics of this
scheduler and develop an offline approximation for it.

\myparab{Observation: Starlink uses an on-satellite controller for scheduling
terminal flows.}
The second peculiar characteristic of the latency measurements from our user
terminals is that within the fifteen-second time interval, latency measurements
the user terminal frequently form parallel bands that are a few milliseconds
apart. 
These bands reflect evidence that radio frames are allocated to user terminals
by an on-satellite controller in a somewhat round-robin fashion.
Further investigation is required to exactly identify the characteristics of
this controller, which we believe to be the on-satellite Medium Access Control
scheduler described in a SpaceX FCC filing \cite{iyer2022system}.

\section{Obtaining Satellite Allocations}
\label{sec:identifying}

Our results show that the Starlink network uses a global 
scheduler to assign satellites to user terminals every 15 seconds 
(\Cref{sec:background}).
Unfortunately, the Starlink mobile app no longer identifies 
the satellite that a user terminal is connected to.
Not having this knowledge limits our ability to reverse-engineer 
the mechanics of Starlink's scheduling algorithms.
In this section, we develop a novel technique that leverages
Starlink's {\em obstruction maps} to identify the satellite allocated 
to a specific user terminal.
%
%
%
At a high-level, our approach involves correlating the publicly 
known positions of the Starlink satellites with observations of connected
satellites recorded by in the obstruction maps of each terminal. 

\begin{figure*}
  \centering 
    \begin{subfigure}[b]{.18\textwidth}
      \includegraphics[width=\textwidth]{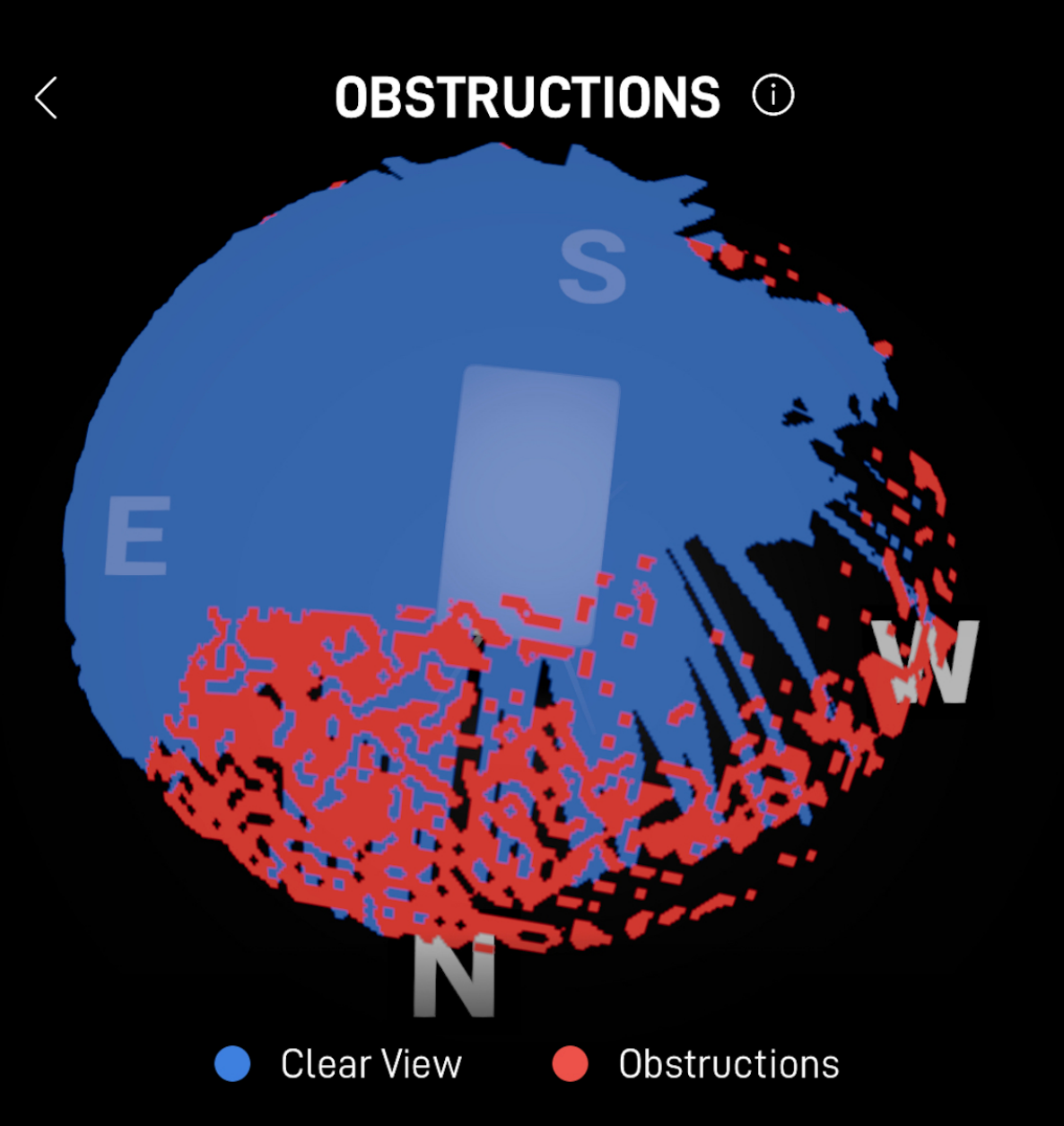}
      \caption{Starlink app.}
      \label{fig:obstruction}
    \end{subfigure}
  \begin{subfigure}[b]{0.19\textwidth}
    \includegraphics[width=\textwidth]{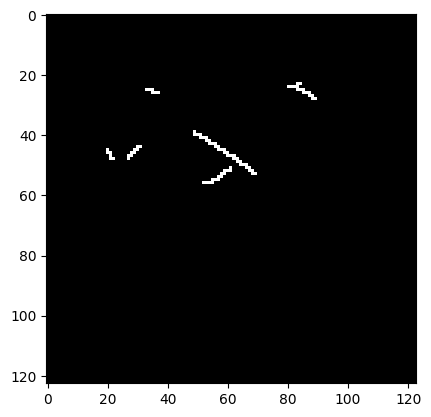}
    \caption{gRPC($x-1$)}
    \label{fig:grpc:1}
  \end{subfigure}
  \begin{subfigure}[b]{0.19\textwidth}
    \includegraphics[width=\textwidth]{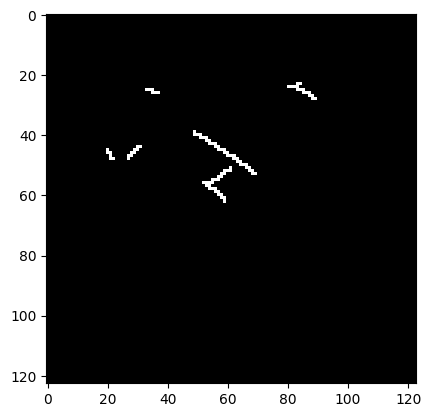}
    \caption{gRPC($x$)}
    \label{fig:grpc:2}
  \end{subfigure}
  \begin{subfigure}[b]{0.19\textwidth}
    \includegraphics[width=\textwidth]{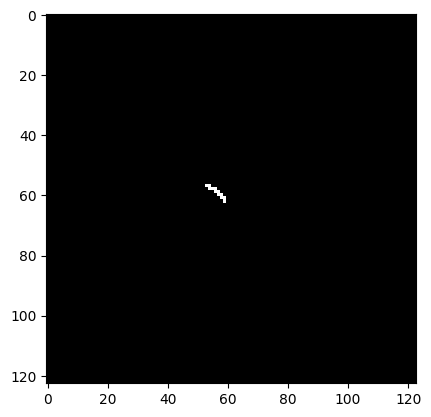}
    \caption{gRPC($x-1$) $\oplus$ gRPC($x$)}
    \label{fig:grpc:xor}
  \end{subfigure}
  \begin{subfigure}[b]{0.18\textwidth}
    \includegraphics[width=\textwidth]{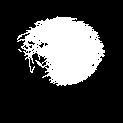}
    \caption{gRPC map after 2 days.}
    \label{fig:grpc:full}
  \end{subfigure}

  \caption{{\bf Obstruction maps} (a) obtained from the Starlink app, 
      (b, c) obtained from gRPC for two consecutive 15-second slots $x-1$ and $x$, 
      (d) their XOR, and 
      (e) the gRPC map after two days without a terminal reset.}
  \label{fig:grpc}
\end{figure*}

\myparab{Data: Obstruction maps.} 
Obstruction maps are 123px x 123px, 2-dimensional images which mark
the trajectory of satellites that recently served the user terminal. 
These images are used to create a 3-dimensional map made available 
to users via the Starlink mobile app. 
The 3-d map is meant to help users identify the quality of the 
location of their terminal, highlighting any physical obstructions 
between their terminal and any satellites meant to serve the 
terminal. \Cref{fig:obstruction} shows an example of this map 
from the Starlink app.
We used {\tt starlink-grpc-tools} \cite{sparky8596:online} to extract 
the 2-d obstruction maps every 15 seconds from each terminal.
The 3-d maps are not possible to obtain programmatically. 
\Cref{fig:grpc:1,fig:grpc:2} shows examples of these maps 
from two consecutive 15-second time slots.
%


\myparab{Data: Satellite positions.}
Positions of Starlink satellites are available publicly 
from a variety of sources in a two-line element(TLE) format. 
We use CelesTrak~\cite{CelesTra44:online} to get the TLEs for 
Starlink satellites.
Since these files only indicate satellite positions every 
six hours, we use the SGP4 satellite propagation 
algorithm~\cite{rhodes2019skyfield} to calculate satellite 
positions, relative to a terminal location, for a
specific point in time. 

%
%
\myparab{Method: Uncovering gRPC obstruction map parameters.}
As seen in \Cref{fig:grpc:1,fig:grpc:2}, the 2-d obstruction maps 
are plain --- only containing white pixels which indicate the 
trajectory of recently connected satellites without any 
further context (\eg the angle of elevation and azimuth).
Identifying these parameters from this map is crucial for 
identifying the connected satellite.
To do this, we align the recorded 2-d maps with the 
3-d maps observed on the Starlink app. From this alignment, 
we find that: 
(1) the 2-d obstruction map is a polar plot centered at 62x62; 
(2) the radius of the polar plot represents the angle of elevation 
and ranges from 25 to 90; and 
(3) the $\theta$ of the polar plot represents the azimuth, 
where $\theta = 0$ represents the North.
Finally, since the obstruction map is a square which contains 
a polar plot, we also need to get the boundaries of the polar 
plot within this square. 
We accomplish this by keeping the terminal online for 
two consecutive days.
In fact, over a 2-day period the terminal will establish connections with satellites from
practically all the regions of the sky that are within the field of view.
In turn, this will result in essentially fully coloring the polar plot region in the gRPC map,
since the gRPC map does not reset (unless the terminal goes offline).
Using this fully colored map (an example is shown in \Cref{fig:grpc:full}), we
find that the radius of the contained polar plot is 45 pixels.

\myparab{Method: Isolating satellite trajectory.}
After recovering the parameters of the gRPC obstruction map, 
we use them to identify the trajectory of the satellite that 
is allocated to the terminal for a specific 15-second slot
(denoted by $x$).
To obtain this trajectory, we perform an XOR operation on the 
obstruction map from $x$ and $x-1$ (\ie the prior 15-second slot).
This will result in the erasure of all satellite trajectories 
which were common to the two figures --- leaving visible only 
the trajectory associated with the satellite connected to the 
terminal during slot $x$.
We note that for this method to work, we require that satellite
trajectories are not overlapping with the trajectories of 
previously connected satellites (since an XOR would erase 
the overlaps). To ensure that this condition is met, we perform 
a terminal reset every 10 minutes (since resetting
the terminal starts a fresh gRPC map).

\myparab{Method: Identifying serving satellite.}
From the prior step, we have the angle of elevation and azimuth 
trajectories of the satellite that served the terminal for a 
specific 15-second slot.
To identify the specific ID of the serving satellite,
we use the computed relative locations of the satellites in the area (as previously explained) to
obtain the IDs of all satellites which are visible to our terminal during the
specific time slot.
Next, we compute the angle of elevations and azimuths for each of these
satellites during the given slot.
Finally, we convert the trajectories to cartesian coordinates and then use the
Dynamic Time Warping (DTW) \cite{muller2007dynamic} distance measure to compute the similarity
of trajectories. 
We select the satellite whose angle of elevation and azimuth trajectories are
the most similar to those recorded from the gRPC maps.
We validate our similarity matching via a manual (visual) pilot test study, in which
the authors manually identified the best match between 500 sets of gRPC and TLE
trajectories. The DTW similarity method and our manual tests overlapped on over
99\% of all outcomes.

\section{Starlink's global scheduler}\label{sec:understanding}
The Starlink network uses a global scheduler to allocate 
user terminals to individual satellites every 15 seconds (\Cref{sec:background}). 
However, the scheduling algorithm is not publicly known. In this 
section, we analyze characteristics of satellites that
were allocated to our terminals with the goal of reverse engineering
the algorithm of the global scheduler.
On average, there are 35--44 satellites in the field of view 
of a user terminal in any 15 second slot. Having identified the 
satellite that a terminal is connected to during a slot 
(\Cref{sec:identifying}), we compare properties of the satellites 
that are selected by the scheduler with those that were available but
not chosen. 

\subsection{Impact of satellite position}

\begin{figure}[]
  \centering
  \includegraphics[width=\linewidth]{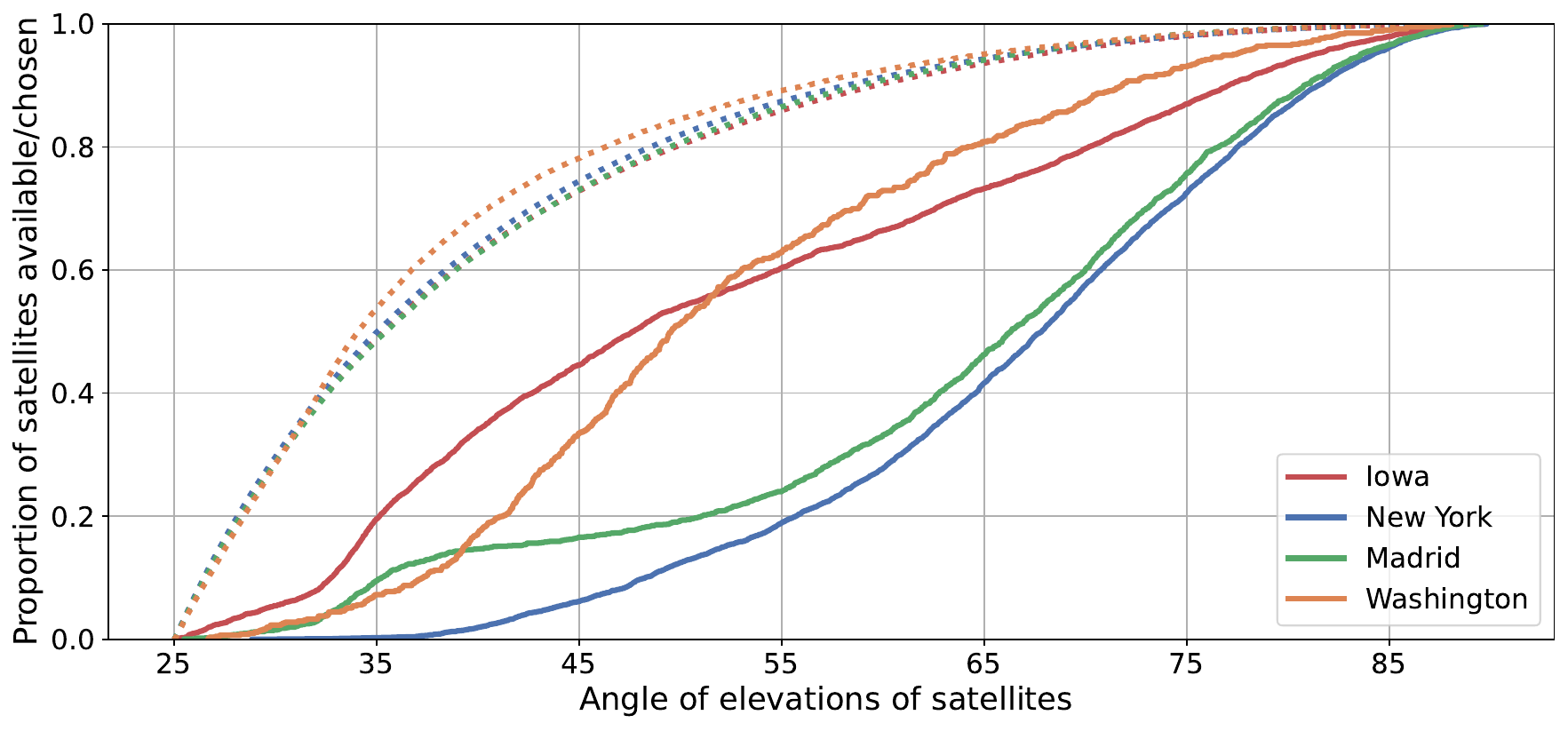}
  \caption{AOEs of available (dotted) vs. selected (solid) satellites.}
  \label{fig:understanding:decision:aoe}
\end{figure}

\myparab{Angle of elevation.}
First, we compare the positions in the sky of satellites that were 
available but not selected to the positions of satellites that were 
selected by the global scheduler. The position of a satellite in the 
sky is defined by its AOE and azimuth with respect to the user 
terminal. \Cref{fig:understanding:decision:aoe} shows that
the median angle of elevation of selected satellites (solid lines) 
is 22.9$^{\circ}$s higher than that of the available but unselected 
satellites (dotted lines). Although only 30\% of all available 
satellites had their AOEs in 45$^{\circ}$ to 90$^{\circ}$ range, 
the global scheduler picked 80\% of satellites from the range
 (averaged over all locations).

\begin{figure}[]
  \centering
  \includegraphics[width=\linewidth]{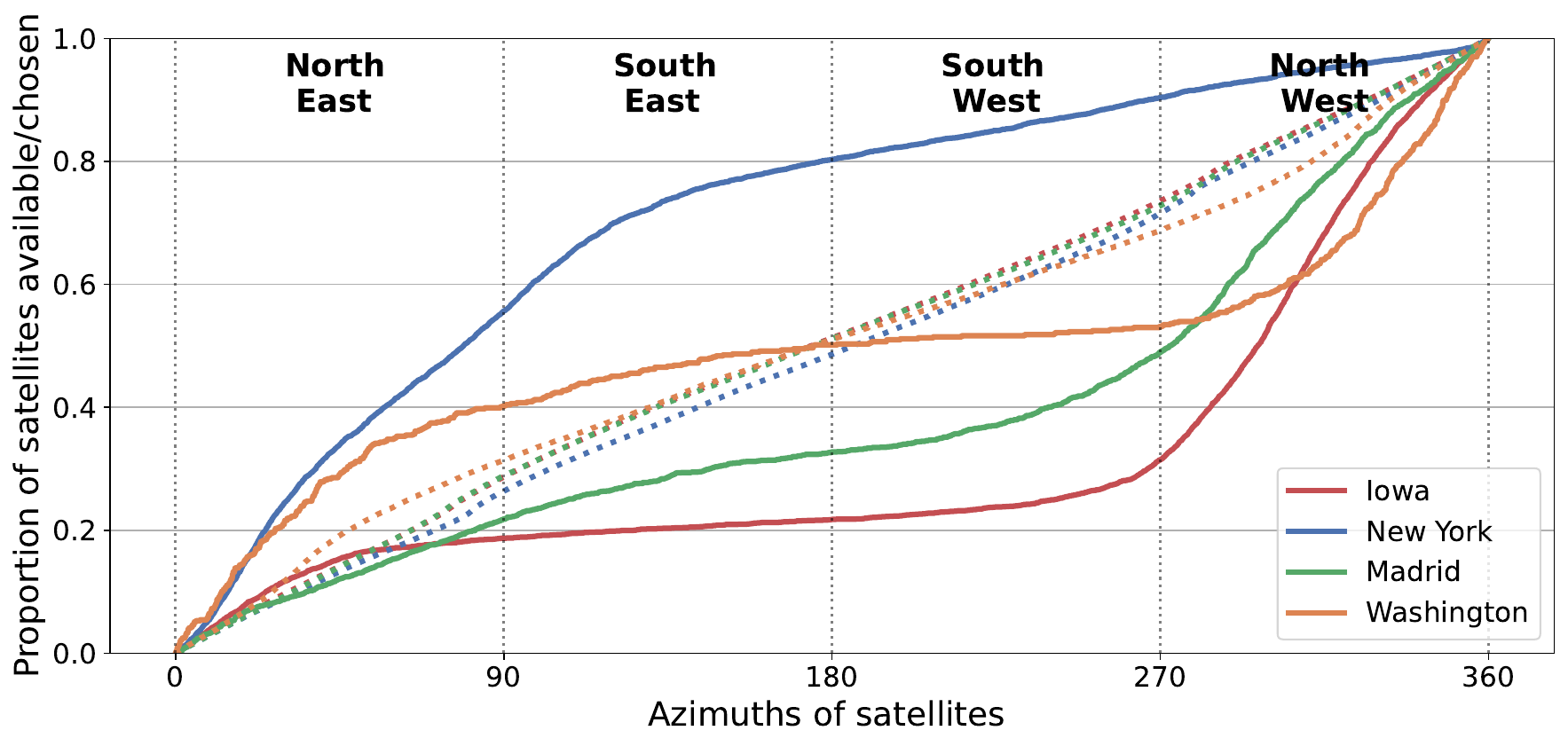}
  \caption{CDF of AZs of satellites available (dotted) vs. AZs of satellites chosen (solid). 
  The x-axis is divided into 4 quadrants of 90 degrees each; the direction of the quadrant relative to the user terminal's face is listed on top of each quadrant.}
  \label{fig:understanding:decision:az}
\end{figure}

\myparab{Direction.}
\Cref{fig:understanding:decision:az} shows the distribution of azimuths 
of the two sets of satellites. The plot is divided into 4 quadrants 
which represent the direction -- stated at the top of each quadrant -- 
of the satellites relative to the face of the user terminal. Although 
the azimuths of available satellites (dotted lines) are evenly 
distributed throughout the 4 quadrants, azimuths of selected satellites
(solid lines) are skewed towards the north of the dish, except 
for the dish in Ithaca, NY. We investigated this difference 
and found that our user terminal in Ithaca was severely obstructed 
by trees towards the north west direction causing it to pick fewer 
satellites from the north west. The terminal in Ithaca was 
assigned only 9.7\% of the satellites from the region compared 
to 55.4\% on average by user terminals in other locations. In other 
locations, 58\% of satellites on average were available towards 
the north of the user terminals, however, the user terminal 
was mapped to satellites from the north 82\% of the times.



\textbf{Rationale:} 
The International Telecommunication Union has
imposed a mandatory geo-stationary orbit exclusion zone, which 
prohibits LEO satellites from transmitting to or receiving from 
a ground station while being in the protected part of the sky~\cite{itulaw}. 
This mandate forces terminals at latitudes more than 
40$^{\circ}$N, the approximate latitude of our terminals,
to point much higher than required by the minimum angle 
of elevation constraint. This is why the global scheduler assigns 
satellites higher up in the fields-of-view of our terminals.
Moreover, satellites with a higher AOE can communicate with 
terminals in a more energy efficient way. The distance between 
the user terminal and satellite increases inversely with AOE. 
As radio frequency (RF) power decreases inversely with distance, 
satellites farther away need to use significantly more power 
to communicate with user terminals. 

\subsection{Impact of satellite launch dates}
\begin{figure}[]
  \centering
  \includegraphics[width=\linewidth]{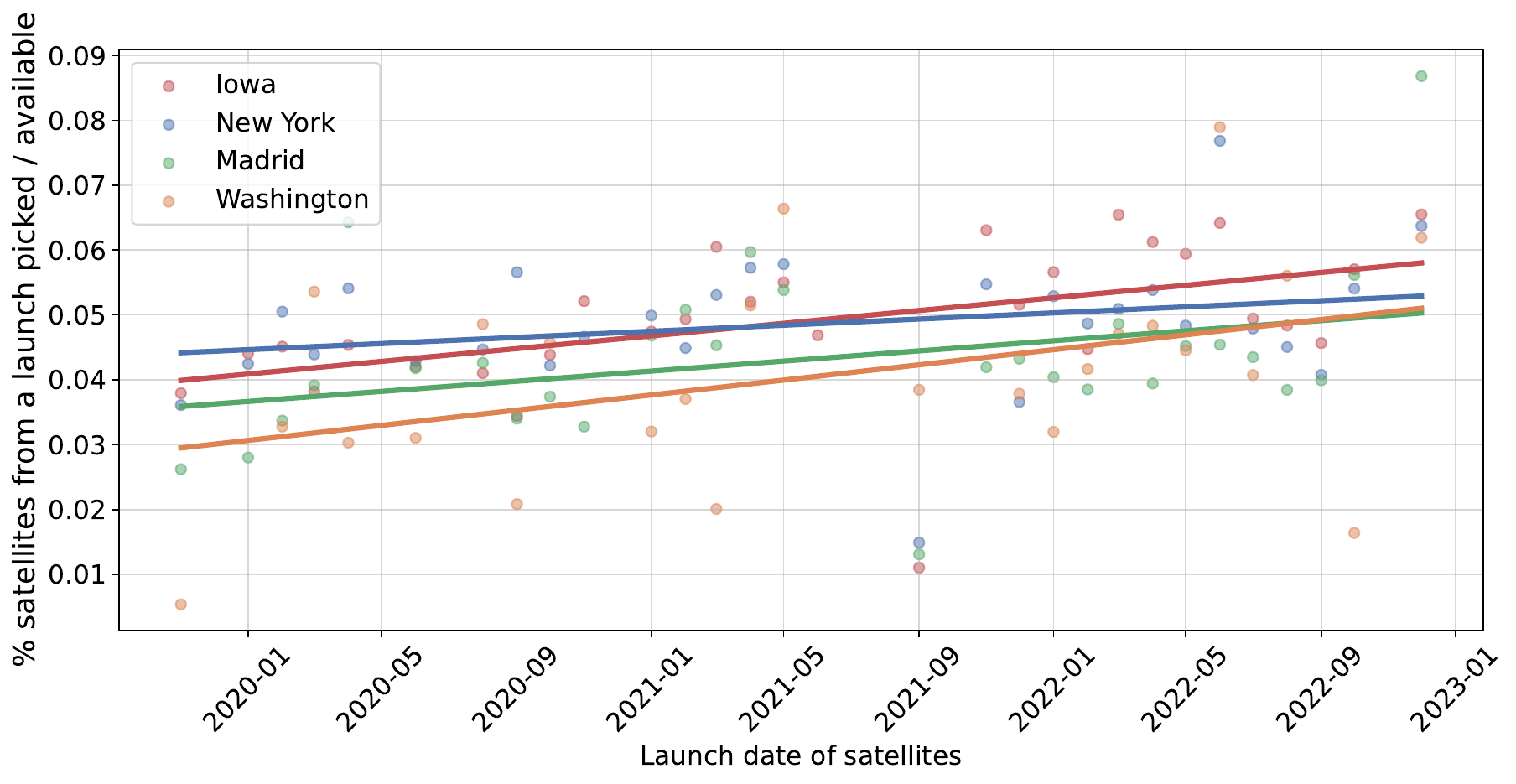}
  \caption{Correlation of probability of picking a satellite from a launch vs. launch date}
  \label{fig:understanding:decision:launch}
\end{figure}

The Starlink constellation consists of more than 4,000 satellites 
which were released in batches since 2018. We analyze whether 
the global scheduler prefers satellites from certain batches over 
others. To achieve this, we bin satellites by the year and 
month of their launch batch. We then compute 
\%(No. of slots a satellite from a launch was 
picked/No. of slots a satellite from a launch was available) for
all 15 second slots in our observation. 
\Cref{fig:understanding:decision:launch} shows the distributions of 
these probabilities with the launch dates of satellites binned 
by year and month. Averaged over all locations, the probability of 
picking a satellite increases with an increase of one month in 
the satellite's launch date.   

\textbf{Rationale: }As Starlink satellites are launched in batches, the difference in service time between the latest and oldest satellites can differ by years. As a Starlink satellite has a service lifetime of about 5 years, some parts of the constellation would be out of service years before others. This would require consistent replacement efforts to maintain the constellation's coverage. Hence, using satellites launched later in time can help offset the discrepancy in service times of satellites allowing the constellation to stay stable for longer.    

\subsection{Impact of being sunlit}

Starlink satellites are equipped with solar panels to provide power 
for their operations. However, as satellites orbit the earth, 
they periodically go in and out of sunlight. To serve user terminals 
during times when a satellite is not under sunlight, it has to 
conserve energy to stay functional until it gets sunlight again. 
For this reason, we analyze whether Starlink's global scheduler 
has preference for sunlit satellites. 

For every 15 second slot, we calculate positions of all available
satellites relative to the sun using the SkyField library. 
We then extract 15 second slots during which at least one sunlit 
and at least one dark satellite was available. During such slots, 
the global scheduler opts for the sunlit satellites $72.3\%$ of 
the time averaged over all locations. We also find that the global 
scheduler only picks dark satellites during 15 second slots where 
the \%(dark/available) satellites is $>= 35\%$ (averaged over all 
locations). Next, we compare the positions of the dark satellites 
that were picked by the global scheduler with the positions of 
their sunlit counterparts. We find that the AOE of dark satellites 
picked by the scheduler was 25 degrees higher than their sunlit 
counterparts. 

\textbf{Rationale: } As mentioned earlier, the RF power decreases 
inversely with distance causing satellites to expend more energy 
in communicating with the user terminal. As dark satellites have 
limited battery to stay functional, the global scheduler assigns 
satellites to user terminals that are higher up in the field-of-view 
to conserve energy.  



\section{Modeling the Global Scheduler}\label{sec:model}

Our analysis has shown that the Starlink global scheduler has specific
preferences --- \ie it selects newer satellites that are sunlit, located towards the North
West, and at a high angle of elevation.
We now use these preferences and our data to build an offline model of the
scheduler.
The goal of this model is to {\em predict} the characteristics of the satellite
allocated to serve a terminal in a given location and time.

\myparab{Feature selection.}
Given a specific location and time, we first identify the satellites that are
available to serve the user terminal using our TLE dataset and the SGP4
algorithm.
Next, we cluster the available satellites based on their azimuth ($\theta$),
angle of elevation ($\phi$), age ($a$), and sunlit status ($\epsilon$) as
follows:
Given a set of satellites ($S$) available at time $t$ for location $l$, the
satellite $s \in S$ with parameters ($\theta_s$, $\phi_s$, $a_s$, $\epsilon_s$)
is placed in the cluster: ($\frac{\theta_s - \mu(\theta)}{\sigma(\theta)}$,
$\frac{\phi_s - \mu(\phi)}{\sigma(\phi)}$,
$\frac{a_s - \mu(a)}{\sigma(a)}$, $\epsilon$). Here $\mu(x)$ and $\sigma(x)$
denote the mean and standard deviation of the feature $x$ (computed from $S$).
Effectively, this approach clusters satellites by how many standard deviations
away from the group mean each of their features are. 
For example, a satellite in the cluster (1, 0, 2, 1) is further than 1
standard deviation away from the mean azimuth and 2 standard deviations away from the mean age
of all the satellites in $S$.
As features for our scheduler, we present the local time ($t_l$) and the count
of satellites available in each cluster at the start of the nearest 15-second
interval.

\myparab{Training, testing, and validating our model.}
Our goal is to construct a model which takes the above features as input and
returns the cluster to which the allocated satellite belongs.
We train a random forest model because of its robustness to overfitting and
the explainability of its predictions.
We got the parameters of this model using grid-search and five-fold
cross-validation. 
We use the data gathered from each location and measured satellite
allocations (\Cref{sec:identifying}) to construct our model --- 80\% of the data
is used to create a training/testing dataset for a five-fold cross-validation
evaluation and the remaining 20\% is used to create a holdout dataset to
validate our model's robustness to overfitting.

\begin{figure}
	\includegraphics[width=0.48\textwidth]{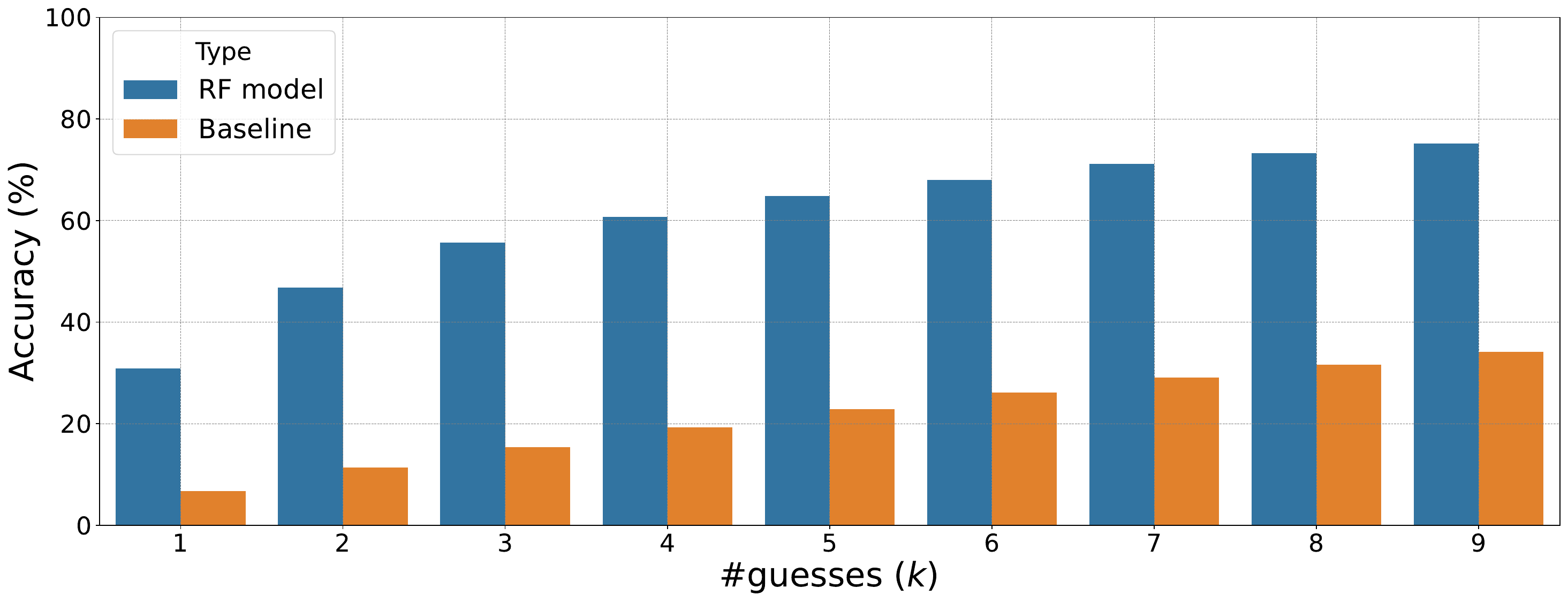}
        \caption{Accuracy of our model compared to baseline using the top-$k$
        accuracy metric.}
	\vspace{-1em}
	\label{fig:model:topk}
\end{figure}

\myparab{Evaluating the model.}
To measure the accuracy of our model, we use a top-k-accuracy metric --- \ie we
analyze the accuracy of the $k$ most likely predictions made by the model.
To put this accuracy in context, we compare it to the baseline model which
simply returns the (top-$k$) cluster(s) with the most number of available
satellites as its prediction.
Our results obtained over our holdout set are shown in \Cref{fig:model:topk}.
The proposed model significantly outperforms the baseline model and predicts the
correct allocated satellite characteristics 65\% of the time ($k$=5 guesses), in
comparison to the baseline of 22\%.

\myparab{Limitations.}
Our model is constructed using only publicly measurable data related to
Starlink's satellites. 
However, based on disclosures in SpaceX's FCC filings \cite{Starlink45:online,iyer2022system}, we expect that
other publicly-unavailable features such as terminal density in a region and
satellite load characteristics will also impact the global scheduler.
Therefore, the performance of our model is constrained by the unavailability of
data. 
Despite this, our model is demonstrably robust to over-fitting and is able to
make reasonably accurate and explainable predictions at a rate far higher than
baseline.

\myparab{Model release.}
In order to facilitate future simulations and evaluations of the Starlink
network, our model will be publicly available on paper acceptance.

\section{Related work}\label{sec:related}
Previous work has studied Starlink's performance 
along the axes of latency and throughput with 
coarse-grained measurements~\cite{uran2021analysis, michel2022first}.
Researchers have found geographic variability \cite{kassem2022browser} 
in Starlink performance. Others have developed simulations
of LEO constellations~\cite{valentine2021developing,kassing2020exploring} 
to suggest improvements to the network~\cite{cakaj2021parameters}.
Researchers have proposed to improve end host performance in 
Starlink using better routing protocols~\cite{handley2019using, 
handley2018delay}, better beamforming techniques \cite{kumar2022dnn}, 
satellite network topology reconfiguration~\cite{bhattacherjee2019network},
clean-slate protocol design~\cite{giuliari2020internet} and 
efficient satellite handoff techniques \cite{juan2022performance, liu2022channel}.
Security researchers have also found potential attack surfaces in Starlink \cite{smailes2023dishing, giuliari2021icarus}.

\section{Conclusions}

In this paper, we leveraged high-frequency and high-fidelity measurements from
four globally distributed vantage points to uncover the presence of a
hierarchical traffic controller --- a global controller for allocating
satellites to user terminals and an on-satellite local controller for scheduling
user flows. 
Using a novel technique for identifying the satellites allocated to terminals,
we identified the characteristics and preferences of the global scheduler and
developed an offline approximation. 
Taken all together, our work is a first step towards understanding Starlink's
traffic engineering decisions and supporting future efforts to model and
evaluate scheduling algorithms for the Starlink network.

\bibliographystyle{ACM-Reference-Format}
\bibliography{reference}


\begin{thebibliography}{25}


\ifx \showCODEN    \undefined \def \showCODEN     #1{\unskip}     \fi
\ifx \showDOI      \undefined \def \showDOI       #1{#1}\fi
\ifx \showISBNx    \undefined \def \showISBNx     #1{\unskip}     \fi
\ifx \showISBNxiii \undefined \def \showISBNxiii  #1{\unskip}     \fi
\ifx \showISSN     \undefined \def \showISSN      #1{\unskip}     \fi
\ifx \showLCCN     \undefined \def \showLCCN      #1{\unskip}     \fi
\ifx \shownote     \undefined \def \shownote      #1{#1}          \fi
\ifx \showarticletitle \undefined \def \showarticletitle #1{#1}   \fi
\ifx \showURL      \undefined \def \showURL       {\relax}        \fi
\providecommand\bibfield[2]{#2}
\providecommand\bibinfo[2]{#2}
\providecommand\natexlab[1]{#1}
\providecommand\showeprint[2][]{arXiv:#2}

\bibitem[\protect\citeauthoryear{??}{itu}{[n. d.]}]%
        {itulaw}
 \bibinfo{year}{[n. d.]}\natexlab{}.
\newblock \bibinfo{title}{{47 CFR § 25.289 - Protection of GSO networks by
  NGSO systems.}}
\newblock
  \bibinfo{howpublished}{\url{https://www.law.cornell.edu/cfr/text/47/25.289}}.
    (\bibinfo{year}{[n. d.]}).
\newblock
\newblock
\shownote{(Accessed on 06/25/2023).}


\bibitem[\protect\citeauthoryear{??}{Cel}{[n. d.]}]%
        {CelesTra44:online}
 \bibinfo{year}{[n. d.]}\natexlab{}.
\newblock \bibinfo{title}{CelesTrak}.
\newblock \bibinfo{howpublished}{\url{https://celestrak.org/}}.
  (\bibinfo{year}{[n. d.]}).
\newblock
\newblock
\shownote{(Accessed on 06/25/2023).}


\bibitem[\protect\citeauthoryear{??}{irt}{[n. d.]}]%
        {irtt:online}
 \bibinfo{year}{[n. d.]}\natexlab{}.
\newblock \bibinfo{title}{IRTT}.
\newblock \bibinfo{howpublished}{\url{https://iperf.fr/}}.
  (\bibinfo{year}{[n. d.]}).
\newblock
\newblock
\shownote{(Accessed on 06/25/2023).}


\bibitem[\protect\citeauthoryear{??}{spa}{[n. d.]}]%
        {sparky8596:online}
 \bibinfo{year}{[n. d.]}\natexlab{}.
\newblock \bibinfo{title}{sparky8512/starlink-grpc-tools: Random scripts and
  other bits for interacting with the SpaceX Starlink user terminal hardware}.
\newblock
  \bibinfo{howpublished}{\url{https://github.com/sparky8512/starlink-grpc-tools}}.
    (\bibinfo{year}{[n. d.]}).
\newblock
\newblock
\shownote{(Accessed on 06/25/2023).}


\bibitem[\protect\citeauthoryear{??}{Sta}{[n. d.]}]%
        {Starlink45:online}
 \bibinfo{year}{[n. d.]}\natexlab{}.
\newblock \bibinfo{title}{Starlink Services LLC Application for ETC Designation
  | PDF Host}.
\newblock
  \bibinfo{howpublished}{\url{https://pdfhost.io/v/BnYWSR~wq_Starlink_Services_LLC_Application_for_ETC_Designation}}.
    (\bibinfo{year}{[n. d.]}).
\newblock
\newblock
\shownote{(Accessed on 06/27/2023).}


\bibitem[\protect\citeauthoryear{Bhattacherjee and Singla}{Bhattacherjee and
  Singla}{2019}]%
        {bhattacherjee2019network}
\bibfield{author}{\bibinfo{person}{Debopam Bhattacherjee} {and}
  \bibinfo{person}{Ankit Singla}.} \bibinfo{year}{2019}\natexlab{}.
\newblock \showarticletitle{Network topology design at 27,000 km/hour}. In
  \bibinfo{booktitle}{{\em Proceedings of the 15th International Conference on
  Emerging Networking Experiments And Technologies}}.
  \bibinfo{pages}{341--354}.
\newblock


\bibitem[\protect\citeauthoryear{Bhosale, Saeed, Bhardwaj, and
  Gavrilovska}{Bhosale et~al\mbox{.}}{2023}]%
        {bhosale2023characterization}
\bibfield{author}{\bibinfo{person}{Vaibhav Bhosale}, \bibinfo{person}{Ahmed
  Saeed}, \bibinfo{person}{Ketan Bhardwaj}, {and} \bibinfo{person}{Ada
  Gavrilovska}.} \bibinfo{year}{2023}\natexlab{}.
\newblock \showarticletitle{A Characterization of Route Variability in LEO
  Satellite Networks}. In \bibinfo{booktitle}{{\em International Conference on
  Passive and Active Network Measurement}}. Springer,
  \bibinfo{pages}{313--342}.
\newblock


\bibitem[\protect\citeauthoryear{Cakaj}{Cakaj}{2021}]%
        {cakaj2021parameters}
\bibfield{author}{\bibinfo{person}{Shkelzen Cakaj}.}
  \bibinfo{year}{2021}\natexlab{}.
\newblock \showarticletitle{The parameters comparison of the “Starlink” LEO
  satellites constellation for different orbital shells}.
\newblock \bibinfo{journal}{{\em Frontiers in Communications and Networks\/}}
  \bibinfo{volume}{2} (\bibinfo{year}{2021}), \bibinfo{pages}{643095}.
\newblock


\bibitem[\protect\citeauthoryear{Giuliari, Ciussani, Perrig, Singla, and
  Zurich}{Giuliari et~al\mbox{.}}{2021}]%
        {giuliari2021icarus}
\bibfield{author}{\bibinfo{person}{Giacomo Giuliari}, \bibinfo{person}{Tommaso
  Ciussani}, \bibinfo{person}{Adrian Perrig}, \bibinfo{person}{Ankit Singla},
  {and} \bibinfo{person}{E Zurich}.} \bibinfo{year}{2021}\natexlab{}.
\newblock \showarticletitle{ICARUS: Attacking low Earth orbit satellite
  networks.}. In \bibinfo{booktitle}{{\em USENIX Annual Technical Conference}}.
  \bibinfo{pages}{317--331}.
\newblock


\bibitem[\protect\citeauthoryear{Giuliari, Klenze, Legner, Basin, Perrig, and
  Singla}{Giuliari et~al\mbox{.}}{2020}]%
        {giuliari2020internet}
\bibfield{author}{\bibinfo{person}{Giacomo Giuliari}, \bibinfo{person}{Tobias
  Klenze}, \bibinfo{person}{Markus Legner}, \bibinfo{person}{David Basin},
  \bibinfo{person}{Adrian Perrig}, {and} \bibinfo{person}{Ankit Singla}.}
  \bibinfo{year}{2020}\natexlab{}.
\newblock \showarticletitle{Internet backbones in space}.
\newblock \bibinfo{journal}{{\em ACM SIGCOMM Computer Communication Review\/}}
  \bibinfo{volume}{50}, \bibinfo{number}{1} (\bibinfo{year}{2020}),
  \bibinfo{pages}{25--37}.
\newblock


\bibitem[\protect\citeauthoryear{Handley}{Handley}{2018}]%
        {handley2018delay}
\bibfield{author}{\bibinfo{person}{Mark Handley}.}
  \bibinfo{year}{2018}\natexlab{}.
\newblock \showarticletitle{Delay is not an option: Low latency routing in
  space}. In \bibinfo{booktitle}{{\em Proceedings of the 17th ACM Workshop on
  Hot Topics in Networks}}. \bibinfo{pages}{85--91}.
\newblock


\bibitem[\protect\citeauthoryear{Handley}{Handley}{2019}]%
        {handley2019using}
\bibfield{author}{\bibinfo{person}{Mark Handley}.}
  \bibinfo{year}{2019}\natexlab{}.
\newblock \showarticletitle{Using ground relays for low-latency wide-area
  routing in megaconstellations}. In \bibinfo{booktitle}{{\em Proceedings of
  the 18th ACM Workshop on Hot Topics in Networks}}. \bibinfo{pages}{125--132}.
\newblock


\bibitem[\protect\citeauthoryear{Hong, Mandal, Al-Fares, Zhu, Alimi, B.,
  Bhagat, Jain, Kaimal, Liang, Mendelev, Padgett, Rabe, Ray, Tewari, Tierney,
  Zahn, Zolla, Ong, and Vahdat}{Hong et~al\mbox{.}}{2018}]%
        {b4andafter}
\bibfield{author}{\bibinfo{person}{Chi-Yao Hong}, \bibinfo{person}{Subhasree
  Mandal}, \bibinfo{person}{Mohammad Al-Fares}, \bibinfo{person}{Min Zhu},
  \bibinfo{person}{Richard Alimi}, \bibinfo{person}{Kondapa~Naidu B.},
  \bibinfo{person}{Chandan Bhagat}, \bibinfo{person}{Sourabh Jain},
  \bibinfo{person}{Jay Kaimal}, \bibinfo{person}{Shiyu Liang},
  \bibinfo{person}{Kirill Mendelev}, \bibinfo{person}{Steve Padgett},
  \bibinfo{person}{Faro Rabe}, \bibinfo{person}{Saikat Ray},
  \bibinfo{person}{Malveeka Tewari}, \bibinfo{person}{Matt Tierney},
  \bibinfo{person}{Monika Zahn}, \bibinfo{person}{Jonathan Zolla},
  \bibinfo{person}{Joon Ong}, {and} \bibinfo{person}{Amin Vahdat}.}
  \bibinfo{year}{2018}\natexlab{}.
\newblock \showarticletitle{B4 and after: Managing Hierarchy, Partitioning, and
  Asymmetry for Availability and Scale in Google's Software-Defined WAN}. In
  \bibinfo{booktitle}{{\em Proceedings of the 2018 Conference of the ACM
  Special Interest Group on Data Communication}} {\em (\bibinfo{series}{SIGCOMM
  '18})}. \bibinfo{publisher}{Association for Computing Machinery},
  \bibinfo{address}{New York, NY, USA}, \bibinfo{pages}{74–87}.
\newblock
\showISBNx{9781450355674}
\showDOI{%
\url{https://doi.org/10.1145/3230543.3230545}}


\bibitem[\protect\citeauthoryear{Iyer, Mahammad, Dandekar, Akella, Chen,
  Barber, and Worters}{Iyer et~al\mbox{.}}{2022}]%
        {iyer2022system}
\bibfield{author}{\bibinfo{person}{Jayasuryan~V Iyer}, \bibinfo{person}{Khasim
  Shaheed~Shaik Mahammad}, \bibinfo{person}{Yashodhan Dandekar},
  \bibinfo{person}{Ramakrishna Akella}, \bibinfo{person}{Chen Chen},
  \bibinfo{person}{Phillip~E Barber}, {and} \bibinfo{person}{Peter~J Worters}.}
  \bibinfo{year}{2022}\natexlab{}.
\newblock \bibinfo{title}{System and method of providing a medium access
  control scheduler}.
\newblock   (\bibinfo{date}{Dec.~27} \bibinfo{year}{2022}).
\newblock
\newblock
\shownote{US Patent 11,540,301.}


\bibitem[\protect\citeauthoryear{Juan, Lauridsen, Wigard, and Mogensen}{Juan
  et~al\mbox{.}}{2022}]%
        {juan2022performance}
\bibfield{author}{\bibinfo{person}{Enric Juan}, \bibinfo{person}{Mads
  Lauridsen}, \bibinfo{person}{Jeroen Wigard}, {and} \bibinfo{person}{Preben
  Mogensen}.} \bibinfo{year}{2022}\natexlab{}.
\newblock \showarticletitle{Performance evaluation of the 5G NR conditional
  handover in LEO-based non-terrestrial networks}. In \bibinfo{booktitle}{{\em
  2022 IEEE Wireless Communications and Networking Conference (WCNC)}}. IEEE,
  \bibinfo{pages}{2488--2493}.
\newblock


\bibitem[\protect\citeauthoryear{Kassem, Raman, Perino, and Sastry}{Kassem
  et~al\mbox{.}}{2022}]%
        {kassem2022browser}
\bibfield{author}{\bibinfo{person}{Mohamed~M Kassem}, \bibinfo{person}{Aravindh
  Raman}, \bibinfo{person}{Diego Perino}, {and} \bibinfo{person}{Nishanth
  Sastry}.} \bibinfo{year}{2022}\natexlab{}.
\newblock \showarticletitle{A browser-side view of starlink connectivity}. In
  \bibinfo{booktitle}{{\em Proceedings of the 22nd ACM Internet Measurement
  Conference}}. \bibinfo{pages}{151--158}.
\newblock


\bibitem[\protect\citeauthoryear{Kassing, Bhattacherjee, {\'A}guas, Saethre,
  and Singla}{Kassing et~al\mbox{.}}{2020}]%
        {kassing2020exploring}
\bibfield{author}{\bibinfo{person}{Simon Kassing}, \bibinfo{person}{Debopam
  Bhattacherjee}, \bibinfo{person}{Andr{\'e}~Baptista {\'A}guas},
  \bibinfo{person}{Jens~Eirik Saethre}, {and} \bibinfo{person}{Ankit Singla}.}
  \bibinfo{year}{2020}\natexlab{}.
\newblock \showarticletitle{Exploring the" Internet from space" with Hypatia}.
  In \bibinfo{booktitle}{{\em Proceedings of the ACM Internet Measurement
  conference}}. \bibinfo{pages}{214--229}.
\newblock


\bibitem[\protect\citeauthoryear{Kumar and Arnon}{Kumar and Arnon}{2022}]%
        {kumar2022dnn}
\bibfield{author}{\bibinfo{person}{Rajnish Kumar} {and} \bibinfo{person}{Shlomi
  Arnon}.} \bibinfo{year}{2022}\natexlab{}.
\newblock \showarticletitle{DNN Beamforming for LEO Satellite Communication at
  Sub-THz Bands}.
\newblock \bibinfo{journal}{{\em Electronics\/}} \bibinfo{volume}{11},
  \bibinfo{number}{23} (\bibinfo{year}{2022}), \bibinfo{pages}{3937}.
\newblock


\bibitem[\protect\citeauthoryear{Liu, Tang, Zhou, Shi, Qian, and Li}{Liu
  et~al\mbox{.}}{2022}]%
        {liu2022channel}
\bibfield{author}{\bibinfo{person}{Yaoqi Liu}, \bibinfo{person}{Xiaogang Tang},
  \bibinfo{person}{Yiqing Zhou}, \bibinfo{person}{Jinglin Shi},
  \bibinfo{person}{Manli Qian}, {and} \bibinfo{person}{Shaoyang Li}.}
  \bibinfo{year}{2022}\natexlab{}.
\newblock \showarticletitle{Channel Reservation based Load Aware Handover for
  LEO Satellite Communications}. In \bibinfo{booktitle}{{\em 2022 IEEE 95th
  Vehicular Technology Conference:(VTC2022-Spring)}}. IEEE,
  \bibinfo{pages}{1--5}.
\newblock


\bibitem[\protect\citeauthoryear{Michel, Trevisan, Giordano, and
  Bonaventure}{Michel et~al\mbox{.}}{2022}]%
        {michel2022first}
\bibfield{author}{\bibinfo{person}{Fran{\c{c}}ois Michel},
  \bibinfo{person}{Martino Trevisan}, \bibinfo{person}{Danilo Giordano}, {and}
  \bibinfo{person}{Olivier Bonaventure}.} \bibinfo{year}{2022}\natexlab{}.
\newblock \showarticletitle{A first look at starlink performance}. In
  \bibinfo{booktitle}{{\em Proceedings of the 22nd ACM Internet Measurement
  Conference}}. \bibinfo{pages}{130--136}.
\newblock


\bibitem[\protect\citeauthoryear{M{\"u}ller}{M{\"u}ller}{2007}]%
        {muller2007dynamic}
\bibfield{author}{\bibinfo{person}{Meinard M{\"u}ller}.}
  \bibinfo{year}{2007}\natexlab{}.
\newblock \showarticletitle{Dynamic time warping}.
\newblock \bibinfo{journal}{{\em Information retrieval for music and motion\/}}
  (\bibinfo{year}{2007}), \bibinfo{pages}{69--84}.
\newblock


\bibitem[\protect\citeauthoryear{Rhodes}{Rhodes}{2019}]%
        {rhodes2019skyfield}
\bibfield{author}{\bibinfo{person}{Brandon Rhodes}.}
  \bibinfo{year}{2019}\natexlab{}.
\newblock \showarticletitle{Skyfield: High precision research-grade positions
  for planets and Earth satellites generator}.
\newblock \bibinfo{journal}{{\em Astrophysics Source Code Library\/}}
  (\bibinfo{year}{2019}), \bibinfo{pages}{ascl--1907}.
\newblock


\bibitem[\protect\citeauthoryear{Smailes, Salkield, Birnbach, Strohmeier, and
  Martinovic}{Smailes et~al\mbox{.}}{2023}]%
        {smailes2023dishing}
\bibfield{author}{\bibinfo{person}{Joshua Smailes}, \bibinfo{person}{Edd
  Salkield}, \bibinfo{person}{Simon Birnbach}, \bibinfo{person}{Martin
  Strohmeier}, {and} \bibinfo{person}{Ivan Martinovic}.}
  \bibinfo{year}{2023}\natexlab{}.
\newblock \showarticletitle{Dishing out DoS: How to Disable and Secure the
  Starlink User Terminal}.
\newblock \bibinfo{journal}{{\em arXiv preprint arXiv:2303.00582\/}}
  (\bibinfo{year}{2023}).
\newblock


\bibitem[\protect\citeauthoryear{Uran, Horvath, and W{\"o}llik}{Uran
  et~al\mbox{.}}{2021}]%
        {uran2021analysis}
\bibfield{author}{\bibinfo{person}{C Uran}, \bibinfo{person}{K Horvath}, {and}
  \bibinfo{person}{H W{\"o}llik}.} \bibinfo{year}{2021}\natexlab{}.
\newblock \bibinfo{booktitle}{{\em Analysis of a Starlink-based Internet
  connection}}.
\newblock \bibinfo{type}{{T}echnical {R}eport}. \bibinfo{institution}{tech.
  rep., Carinthia University of Applied Science, Research Group ROADMAP-5G}.
\newblock


\bibitem[\protect\citeauthoryear{Valentine and Parisis}{Valentine and
  Parisis}{2021}]%
        {valentine2021developing}
\bibfield{author}{\bibinfo{person}{Aiden Valentine} {and}
  \bibinfo{person}{George Parisis}.} \bibinfo{year}{2021}\natexlab{}.
\newblock \showarticletitle{Developing and experimenting with leo satellite
  constellations in omnet++}.
\newblock \bibinfo{journal}{{\em arXiv preprint arXiv:2109.12046\/}}
  (\bibinfo{year}{2021}).
\newblock


\end{thebibliography}

\end{document}